\documentclass[aps,preprintnumbers,superscriptaddress,showpacs]{revtex4}
\usepackage{graphicx}
\usepackage{amsmath}
\usepackage{amssymb}
\usepackage{bm}

\begin{document}

\title{Which scale breaks quarkonium? A controlled comparison of
density and momentum relaxation in a bottom-up
Einstein-Maxwell-axion model}

\author{Wenxing Cheng}
\email{chengwenxing@cug.edu.cn} \affiliation{School of Mathematics
and Physics, China University of Geosciences, Wuhan 430074, China}

\author{Yunxianghui Wang}
\email{wyxh75891215@163.com} \affiliation{School of Mathematics
and Physics, China University of Geosciences, Wuhan 430074, China}

\author{Zi-qiang Zhang}
\email{zhangzq@cug.edu.cn} \affiliation{School of Mathematics and
Physics, China University of Geosciences, Wuhan 430074, China}

\begin{abstract}
In a strongly coupled medium that is both dense and dissipative,
static color screening and thermodynamic response need not be
controlled by the same scale. We study this question in a
holographic Einstein-Maxwell-axion model where a phenomenological
density parameter $\mu$ and a translational symmetry breaking
parameter $\beta$ are independent controls. We use this model as a
controlled geometric laboratory rather than a quantitative QCD
description: it has no dilaton or non-Abelian flavor fields, and
$\mu$ and $\beta$ are phenomenological controls rather than QCD
chemical potentials. At fixed temperature we compute the static
potential, the probe pair free energy, and the running coupling.
At zero temperature ($T=0$) all quantities are expressed in units
set by powers of $z_h$. We define finite difference logarithmic
sensitivities and scan the admissible parameter space. In the
scanned region, $\mu$ affects the potential and running coupling
more strongly, while $\beta$ affects the probe pair free energy
more strongly. At $T=0$ the horizon remains finite for nonzero
$\mu$ or $\beta$, so the horizon entropy density does not vanish,
and the renormalized free energy differs from the potential by a
term proportional to $1/z_h$. The hierarchy persists in the
dissociation scale and is more pronounced at zero temperature. It
follows from the different $z$ dependence of the $\mu^2 z^4$ and
$\beta^2$ terms in the metric. This pattern is a model level
diagnostic for more realistic backgrounds with a dilaton and
non-Abelian flavor dynamics.
\end{abstract}


\maketitle

\section{Introduction}
Heavy quarks are among the few probes whose in-medium modification
can be computed on both the lattice and the weak-coupling side,
and their suppression patterns are central to the interpretation
of quarkonium yields at the Relativistic Heavy Ion Collider (RHIC)
and the Large Hadron Collider (LHC) \cite{EVS}. The gauge/gravity
duality
\cite{Maldacena:1997re,Witten:1998qj,Gubser:1998bc,MaldacenaReview}
has become a standard nonperturbative tool for the strongly
coupled regime, with a broad literature on transport,
thermodynamics, and heavy quark dynamics \cite{JCA,OD0,JSa}. In
this framework a static quark-antiquark ($Q\bar{Q}$) pair on the
boundary is dual to an open string in the bulk \cite{FW4}, and the
Nambu-Goto action for the associated Wilson loop yields the static
potential, the probe pair free energy, and the running coupling
\cite{HPW3,HPW4,FW5,SH,SH1,HPW,HPW1,HPW2,FW1,FW2,FW3,RW,DW,CO,EB,hc1,hc2,hc3,hc4,hc5,hc6,hc7,hc8,hc9,hc10,hc11,hc12,hc13,hc14,hc15}.

A strongly coupled medium can be both dense and dissipative. In
the bottom-up model considered here, finite density is encoded by
a phenomenological chemical potential $\mu$, while spatial
inhomogeneities that relax momentum are encoded holographically by
a translational symmetry breaking (TSB) scale $\beta$. Several TSB
constructions have been developed
\cite{TSB1,TSB2,TSB3,TSB4,TSBmajor,TSBmajor1}, and many
holographic studies have shown that background parameters can
significantly alter color screening and quarkonium stability
\cite{CWX,HP1,HP2,HP3,HP4,HP5,HP6,HP8,HP9,HP10,HP11,HP12,HP13,HP14,Q1,Q2,AH1,AH2,R1,R6,DC,CX2,FS,DW1,SM,qx1,qx2,qx3,qx4,qx5,qx6,qx7,qx8,qx9,qx10,qx11}.
However, a remaining question is whether the two scales act on the
same observables in the same way. This bears on whether the static
color force and the thermodynamic response of a strongly coupled
medium are controlled by the same underlying physics.

This question cannot be answered within the backgrounds that have
dominated the literature, because each of them isolates only one
scale. Charged Anti-de Sitter (AdS) backgrounds with $\beta=0$
\cite{CWX,hc10} isolate density but eliminate momentum relaxation;
by contrast, neutral TSB backgrounds with $\mu=0$ isolate
dissipation but eliminate density. When both scales are present,
they deform the same bulk geometry, and a scan over parameter
values alone does not reveal which deformation is responsible for
a given shift. Therefore, a background is required in which $\mu$
and $\beta$ are independent controls, along with observables that
probe different regions of that background.

The Einstein-Maxwell-axion (EMa) geometry of Ref.~\cite{TSBmajor1}
provides such control. In particular, the two parameters enter the
metric function differently: $\mu$ contributes a term $\mu^2
z^4/(3z_h^4)$ that grows toward the horizon and produces a finite
Maxwell flux in the near horizon region, while $\beta$ contributes
a constant $-\beta^2/4$ that shifts the horizon condition and the
entropy but does not produce a localized near horizon deformation.
Earlier studies of this background have focused on dynamical
probes: drag forces \cite{sara} and the Schwinger effect
\cite{sara1}. The latter has been analyzed with both $\mu$ and
$\beta$ present \cite{sara1}, but for pair production rather than
for the static $Q\bar{Q}$ observables considered here.

As a bottom-up construction, the EMa model contains neither a
dilaton field with a potential informed by QCD nor a non-Abelian
bulk Yang-Mills sector implementing the flavor symmetry.
Consequently, it does not reproduce the lattice QCD equation of
state at the quantitative level, and its single Abelian chemical
potential is not identified with $\mu_B$, $\mu_Q$, or $\mu_S$. In
particular, the second derivative of the pressure with respect to
$\mu$ in this model is not the lattice baryon susceptibility. The
model is therefore not equivalent to ``holographic QCD'' and does
not claim direct quantitative applicability to real-world QCD. The
purpose of this work is a controlled model level comparison: in a
background where density and momentum relaxation are independent
geometric controls, which one dominates which observable, and why?

We compute the static potential, the probe pair free energy, and
the running coupling at a fixed finite temperature, together with
their zero temperature counterparts. For convenience in the
numerical analysis, we introduce dimensionless quantities at
finite temperature, while at zero temperature all quantities are
expressed in units set by the appropriate powers of $z_h$. To
avoid relying solely on representative curves, we define
dimensionless finite difference logarithmic sensitivities and
mixed logarithmic finite differences, and we scan the admissible
parameter space. The main finding is a hierarchy in the scanned
region: $\mu$ controls the potential and the running coupling,
while $\beta$ controls the probe pair free energy. At $T=0$, where
the horizon remains finite, the horizon entropy density does not
vanish and the renormalized free energy is related to the
potential by $V^{T=0}=F^{T=0}+1/(\pi\alpha' z_h)$. Compared with
the finite temperature results, the dominance of $\mu$ in the
potential and that of $\beta$ in the probe pair free energy appear
more pronounced at zero temperature. The geometric reason for the
hierarchy is the different $z$ dependence of the two terms: $\mu$
deforms the near horizon region that governs string screening,
whereas $\beta$ modifies the geometry more globally. Because this
is a feature of a specific bottom-up geometry, it is formulated as
a model level diagnostic that can be tested in more realistic
constructions with a dilaton and non-Abelian flavor fields. A
model that violates the pattern may possess a qualitatively
different matter sector or symmetry breaking mechanism.

The paper is organized as follows. Sec.~\ref{sec:setup} sets up
the EMa geometry and the string calculation, including the
renormalization scheme and the distinction between the potential
and the probe pair free energy. Sec.~\ref{sec:finiteT} presents
the finite temperature comparison and the sensitivity analysis.
Sec.~\ref{sec:zeroT} discusses the zero temperature limit with the
finite horizon and the corrected $V$--$F$ relation.
Sec.~\ref{sec:summary} summarizes the hierarchy, its limitations,
and the model level diagnostic it implies.

\section{Heavy quark observables in the EMa background}\label{sec:setup}
\subsection{EMa background geometry}

We use the EMa action of Ref.~\cite{TSBmajor1},
\begin{equation}
    S_0=\int_{M} \sqrt{-g}
\left[R-2\Lambda-\frac{1}{2}\sum_{I=1}^{d-1}\big(\partial
\psi_I\big)^2- \frac{1}{4}F_{\mu\nu}F^{\mu\nu}\right]d^{d+1}x-2
\int_{\partial M} \sqrt{-\gamma}\, K\, d^{d}x,
\end{equation}
with $\Lambda=-d(d-1)/(2l^2)$, $16\pi G=1$, and $l=1$. To be
consistent with the five-dimensional background, we consider the
case in which the theory is coupled to three neutral, massless
scalar fields ($d=4$).
 By deriving the equations of motion from the
bulk action and introducing the coordinate transformation $z=1/r$,
we obtain the following static, homogeneous, and isotropic metric
\begin{equation}
    ds^2=-f(z)dt^2+\frac{d\vec{x}^2}{z^2}+\frac{dz^2}{z^4f(z)},\label{metric}
\end{equation}
where
\begin{equation}
f(z)=\frac{1}{z^2}-\frac{\beta^2}{4}-m_0 z^2+\frac{\mu^2
z^4}{3z_h^4},\qquad \label{f}A_t(z)=\mu(1-\frac{z^2}{z_h^2}).
\end{equation}
The charge density is $\rho=2\mu/z_h^2$ in our normalization. The
horizon condition $f(z_h)=0$ fixes
\begin{equation}
    m_0=\frac{1}{z_h^4}(1+\frac{\mu^2 z_h^2}{3}-\frac{\beta^2 z_h^2}{4}).\label{m}
\end{equation}

It is convenient to introduce the dimensionless blackening
function $h(z)=z^2 f(z)$. One obtains
\begin{equation}
h(z)=1-\frac{z^4}{z_h^4}-\frac{\beta^2
z^2}{4}(1-\frac{z^2}{z_h^2})-\frac{\mu^2
z^4}{3z_h^2}(1-\frac{z^2}{z_h^2}). \label{h}
\end{equation}

In these variables $ds^2=z^{-2}[-hdt^2+d\vec{x}^{\,2}+dz^2/h]$.
For $\beta=0$ the solution reduces to a planar Reissner-Nordstr?m
Anti-de Sitter (RN-AdS) black brane; for $\mu=0$ it reduces to a
neutral Einstein-axion black brane; for $\mu=\beta=0$ it reduces
to AdS-Schwarzschild. These limits distinguish the two physical
sources without changing the model.

The Hawking temperature, obtained from the surface gravity at the
horizon \cite{ARC}, is
\begin{equation}
T=-\frac{z_h^2 f'(z_h)}{4\pi} =\frac{1}{\pi z_h} (1-\frac{\beta^2
z_h^2}{8}-\frac{\mu^2 z_h^2}{6}). \label{T}
\end{equation}
Throughout, we fix $T$ and vary $\mu$ and $\beta$ separately, so
that their effects are not conflated through the temperature.
Here, $\mu$ is a phenomenological density parameter of this
bottom-up model and is not identified with any QCD chemical
potential; $\beta$ is a momentum relaxation scale.

\subsection{Heavy quark potential from the Wilson loop}
The static potential follows from the rectangular Wilson loop
\cite{V0,V1,V2},
\begin{equation}
\langle W(C)\rangle=e^{-i\mathcal{T}V_{(\beta,\mu)}},
\end{equation}
with $\langle W(C)\rangle\sim e^{iS_\mathrm{NG}}$ and
$V_{(\beta,\mu)}=-S_\mathrm{NG}/\mathcal{T}$. For a string
parametrized by $t=\tau$, $x_1=\sigma$, $z=z(\sigma)$, the
Nambu-Goto action is
\begin{equation}
S_\mathrm{NG}=-\frac{1}{2\pi\alpha'}\int d\tau d\sigma\,
\sqrt{-\det g_{ab}},
\end{equation}
and the Lagrangian density is
$\mathcal{L}=\sqrt{a(z)+b(z)\dot{z}^2}$ with $a(z)=f(z)/z^2$ and
$b(z)=1/z^4$. The conserved Hamiltonian gives
\begin{equation}
\dot{z}=\sqrt{\frac{a^2(z)-a(z)a(z_c)}{a(z_c)b(z)}},\label{dotz}
\end{equation}
and the interquark distance is
\begin{equation}
L=2\int_{0}^{z_c}dz\sqrt{\frac{a(z_c)b(z)}{a^2(z)-a(z)a(z_c)}}.\label{L}
\end{equation}
After subtracting the self-energy of two isolated quarks,
\begin{equation}
V_{(\beta,\mu)}=\frac{1}{\pi\alpha'}\int_{0}^{z_c}
dz\left[\sqrt{\frac{a(z)b(z)}{a(z)-a(z_c)}}-\sqrt{b(z)}\right]-\frac{1}{\pi\alpha'}
\int_{z_c}^{z_h}dz\sqrt{b(z)}.\label{V0}
\end{equation}

\subsection{Free energy and running coupling}
To further investigate the properties of heavy quarks in the
strongly coupled medium, we consider the free energy of a static
heavy quark-antiquark pair. The probe pair free energy follows
from the same Wilson-loop correlator,
\begin{equation}
F_{(\beta,\mu)}\sim-\frac{S_\mathrm{NG}}{\mathcal{T}},\qquad
\mathcal{T}\rightarrow\infty,
\end{equation}
with the minimal subtraction of Refs.~\cite{CO,EB}. The
renormalized expression is
\begin{equation}
F_{(\beta,\mu)}=\frac{1}{\pi\alpha'}\int_{0}^{z_c}dz\left[\sqrt{\frac{a(z)b(z)}{a(z)-a(z_c)}}
-\frac{1}{z^2}\right]-\frac{1}{\pi\alpha'}\frac{1}{z_c}.\label{Fc}
\end{equation}
The quantity $F_{(\beta,\mu)}$ characterizes the change in the
thermodynamic potential associated with introducing a probe
$Q\bar{Q}$ pair into a strongly coupled medium. Unlike the
thermodynamic potential of the background medium as a whole, this
quantity reflects the response of the medium to the $Q\bar{Q}$
pair probes. Thus, the background thermodynamic relation $F=U-TS$
is not used to interpret $F_{(\beta,\mu)}$ directly. In the short
distance limit, medium effects become negligible, and the heavy
quark free energy approaches the Coulombic potential in vacuum.

The $V_{(\beta,\mu)}$ defined in the preceding subsection can be
understood as the free-energy difference between an interacting
heavy quark-antiquark pair and two isolated heavy quarks $F_Q$.
Therefore, the heavy quark potential primarily characterizes the
relative binding properties of the heavy quark-antiquark pair,
whereas the heavy quark free energy additionally incorporates the
free-energy contributions associated with the interactions of
individual heavy quarks with the background medium. Consequently,
the relation between the heavy quark potential and the free energy
can be expressed as

\begin{equation}
V_{(\beta,\mu)}=F(\beta,\mu)-2F_Q=F_{(\beta,\mu)}+\frac{1}{\pi\alpha'
z_h}. \label{VFrelation}
\end{equation}
At finite temperature this difference is nonzero; at $T=0$ it
remains nonzero whenever $z_h$ is finite, as discussed in
Sec.~\ref{sec:zeroT}. This relation is used below for the zero
temperature limit.

The running coupling is defined through \cite{RW}
\begin{equation}
\alpha_{(\beta,\mu)}=\frac{3}{4}L^2
\frac{dF_{(\beta,\mu)}}{dL}.\label{a}
\end{equation}
Because the relation between $L$ and $z_c$ is implicit and
nonlinear, all results below are obtained by numerical evaluation.

\section{Numerical results at finite temperature}\label{sec:finiteT}
\subsection{Parameter constraints and numerical setup}
Positivity of $T$ requires
\begin{equation}
 \frac{24}{z_h^2}-3\beta^2-4\mu^2>0,
 \end{equation}
and the null energy condition requires
\begin{equation}
\beta^2+\frac{4z^4\mu^2}{z_h ^4}\geq0.
\end{equation}
We rescale by $2\pi T$,
\begin{equation}
\beta_1=\frac{\beta}{2\pi T}, \qquad \mu_1=\frac{\mu}{2\pi T},
\qquad \frac{1}{z_{h1}}=\frac{1}{z_h 2\pi T}, \qquad
\frac{1}{z_1}=\frac{1}{z 2\pi T}, \label{eq:rescale}
\end{equation}
and drop the subscripts below. The rescaled horizon condition is
\begin{equation}
\frac{2}{z_h}-\frac{\beta^2 z_h}{4}-\frac{\mu^2 z_h}{3}=1.
\end{equation}

For convenience in the subsequent numerical analysis, we employ
the rescaling defined in \eqref{eq:rescale} only at finite
temperature to introduce the following dimensionless quantities:
\begin{equation}
\frac{V(\beta,\mu)}{2\pi T\sqrt{\lambda}}
=\frac{1}{\pi}\int_0^{z_c}dz
\left[\sqrt{\frac{a(z)b(z)}{a(z)-a(z_c)}} -\sqrt{b(z)}\right]
-\frac{1}{\pi}\int_{z_c}^{z_h}dz\sqrt{b(z)}. \label{V1}
\end{equation}

\begin{equation}
\frac{F(\beta,\mu)}{2\pi T\sqrt{\lambda}}
=\frac{1}{\pi}\int_0^{z_c}dz
\left[\sqrt{\frac{a(z)b(z)}{a(z)-a(z_c)}} -\frac{1}{z^2}\right]
-\frac{1}{\pi}\frac{1}{z_c}. \label{F1}
\end{equation}

\begin{equation}
\frac{\alpha_{(\beta,\mu)}}{\sqrt{\lambda}} =\frac{3}{4}(2\pi
TL)^2 \frac{d\left(F_{(\beta,\mu)}/(2\pi T\sqrt{\lambda})\right)}
{d(2\pi TL)}. \label{alpha1}
\end{equation}

\begin{equation}
2\pi TL=2\int_0^{z_c}dz \sqrt{\frac{a(z_c)b(z)}
{a^2(z)-a(z)a(z_c)}}. \label{L1}
\end{equation}
where we have used the relation $\sqrt{\lambda}=1/\alpha'$, with
$\lambda$ denoting the 't Hooft coupling and the AdS radius $R$
set to unity. Here, $z$ denotes the rescaled radial coordinate
$z_1$, with the subscript omitted for simplicity.

Integrals are evaluated with adaptive quadrature at relative error
below $10^{-8}$. We use two scans: fixed $\beta$ with varying
$\mu$, and fixed $\mu$ with varying $\beta$, so that the
comparison is made at matched parameter points. The representative
scans shown in the figures use $\mu,\beta=1,1.5,2$; the
sensitivity analysis is performed on the same grid.

\subsection{Sensitivity measures}
To move beyond a qualitative comparison of a few curves, we define
dimensionless finite difference logarithmic sensitivities for an
observable $Y$,
\begin{equation}
\chi_\mu^Y\equiv \frac{\Delta\ln |Y|}{\Delta\ln\mu}, \qquad
\chi_\beta^Y\equiv \frac{\Delta\ln |Y|}{\Delta\ln\beta},
\label{chi}
\end{equation}
and mixed logarithmic finite differences
\begin{equation}
\chi_{\mu\beta}^Y\equiv \frac{\Delta_\mu\Delta_\beta\ln |Y|}
{\Delta\ln\mu\,\Delta\ln\beta}. \label{mixed}
\end{equation}

\begin{figure}
\centering
\includegraphics[width=0.48\textwidth]{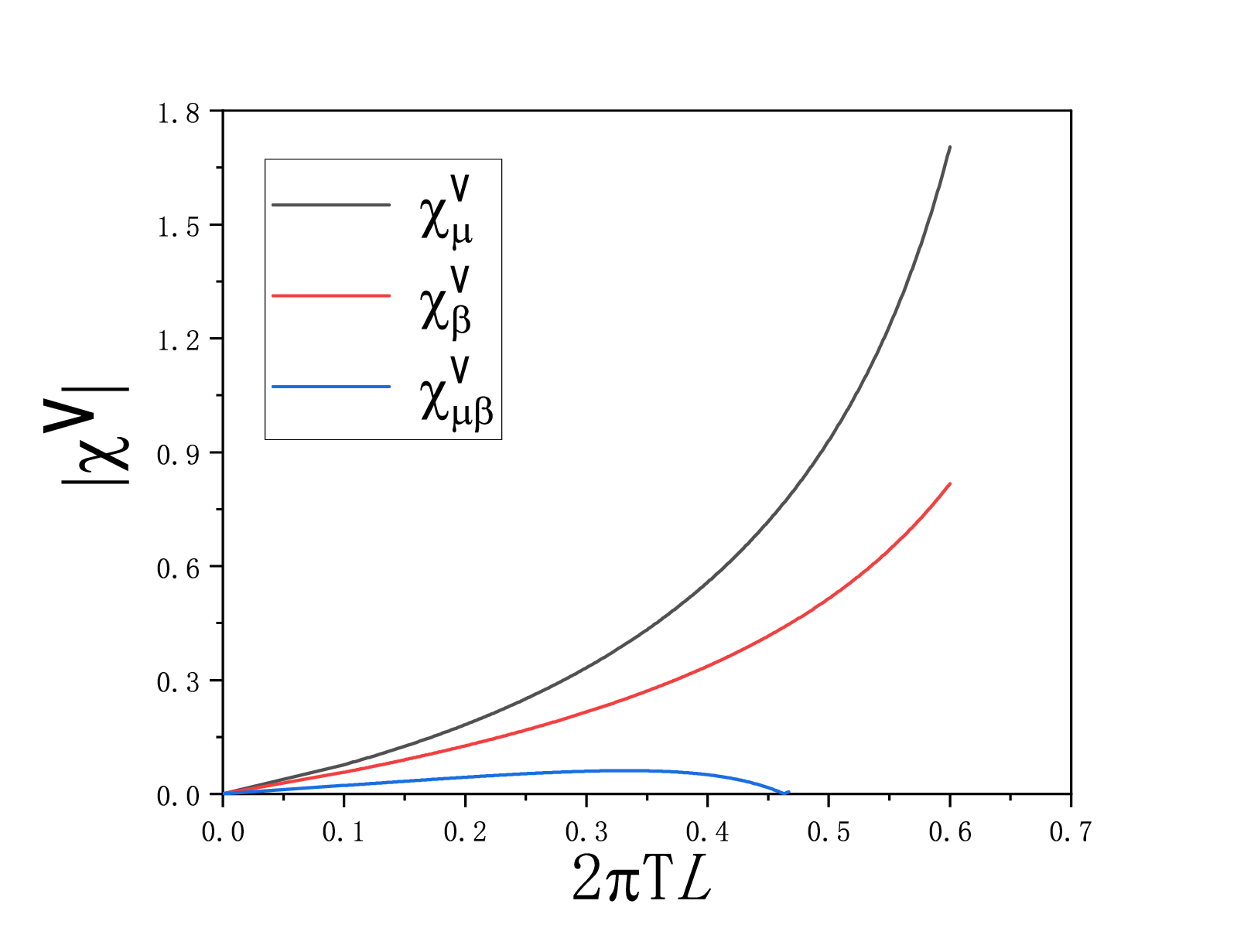}\hfill
\includegraphics[width=0.48\textwidth]{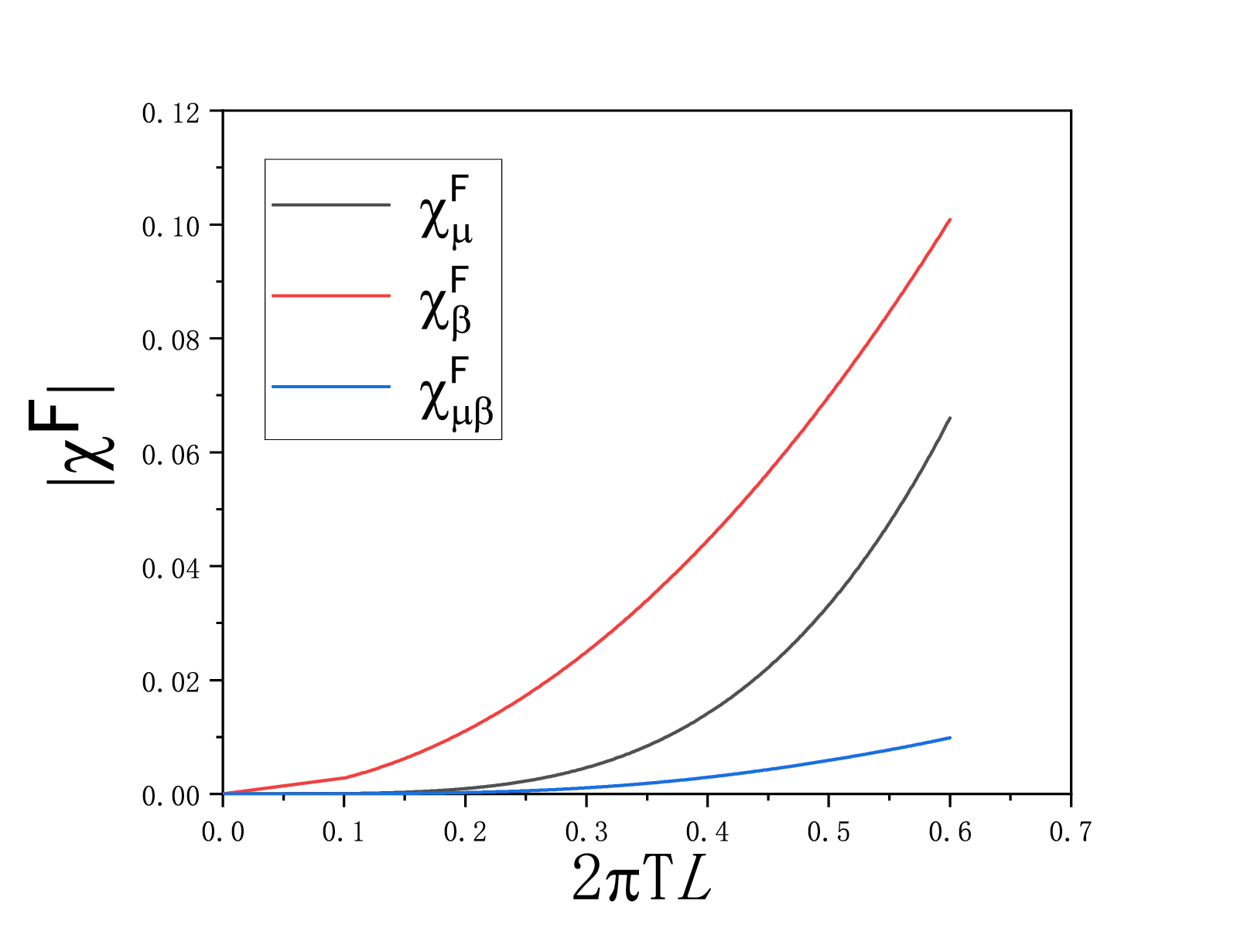}
\caption{Finite difference logarithmic sensitivities at finite
temperature. Left: $|\chi_\mu^V|$, $|\chi_\beta^V|$, and
$|\chi_{\mu\beta}^V|$ versus $2\pi TL$. Right: $|\chi_\mu^F|$,
$|\chi_\beta^F|$, and $|\chi_{\mu\beta}^F|$ versus $2\pi TL$. In
both panels, $\mu$ varies from 1 to 2 with fixed $\beta=1$, and
$\beta$ varies from 1 to 2 with fixed $\mu=1$. The mixed
logarithmic finite differences are evaluated using
$(\mu,\beta)=(1,1),(2,1),(1,2),(2,2)$. All quantities are
dimensionless.} \label{fig:sensitivity}
\end{figure}

Here $Y$ can be the screening length $L_s$, the potential at a
fixed separation $V(L_0)$, the probe pair free energy at a fixed
separation $F(L_0)$, or the running coupling at a fixed separation
$\alpha(L_0)$. The screening length $L_s$ is defined as the
largest $L$ for which the potential remains negative. We evaluate
these finite differences numerically using the parameter values
shown in the plots. In particular, for the potential and the probe
pair free energy, we compare the effects of varying $\mu$ and
$\beta$ from 1 to 2 in the left and right panels of
Figs.~\ref{fig:sensitivity} respectively, at the same interquark
separation and temperature. The purpose is to test whether the
inequalities
\begin{equation}
|\chi_\mu^{V}|>|\chi_\beta^{V}|, \qquad
|\chi_\beta^{F}|>|\chi_\mu^{F}| \label{ineq}
\end{equation}
hold throughout the scanned separation interval. As shown in
Fig.~\ref{fig:sensitivity}, the inequalities are satisfied
throughout this interval. Specifically, the left panel of
Fig.~\ref{fig:sensitivity} shows that
$|\chi_\mu^V|>|\chi_\beta^V|>|\chi_{\mu\beta}^V|$ over the
separation range in which all three curves are shown, indicating
that the chemical potential has a stronger effect on the heavy
quark potential than the translational symmetry breaking
parameter. The mixed logarithmic finite difference first increases
and then decreases, approaching zero near $2\pi TL\approx0.47$. In
contrast, the right panel of Fig.~\ref{fig:sensitivity} shows a
different ordering,
$|\chi_\beta^F|>|\chi_\mu^F|>|\chi_{\mu\beta}^F|$, indicating that
the translational symmetry breaking parameter has a stronger
effect on the probe pair free energy than the chemical potential.
Although the mixed logarithmic finite difference increases
gradually with the interquark separation, it remains subleading
throughout the displayed separation interval.

\subsection{Potential and screening}
\begin{figure}
\centering
\includegraphics[width=0.48\textwidth]{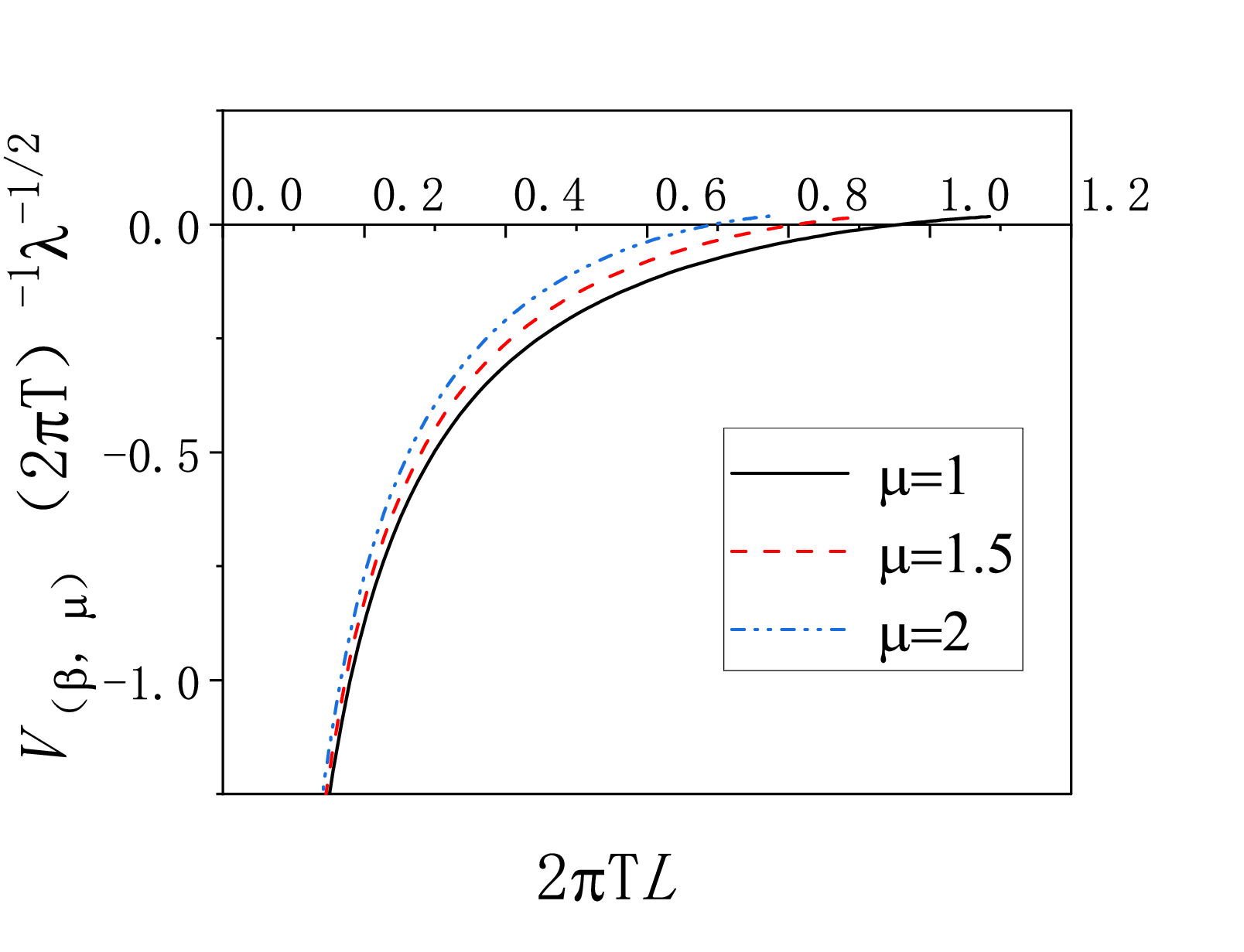}\hfill
\includegraphics[width=0.48\textwidth]{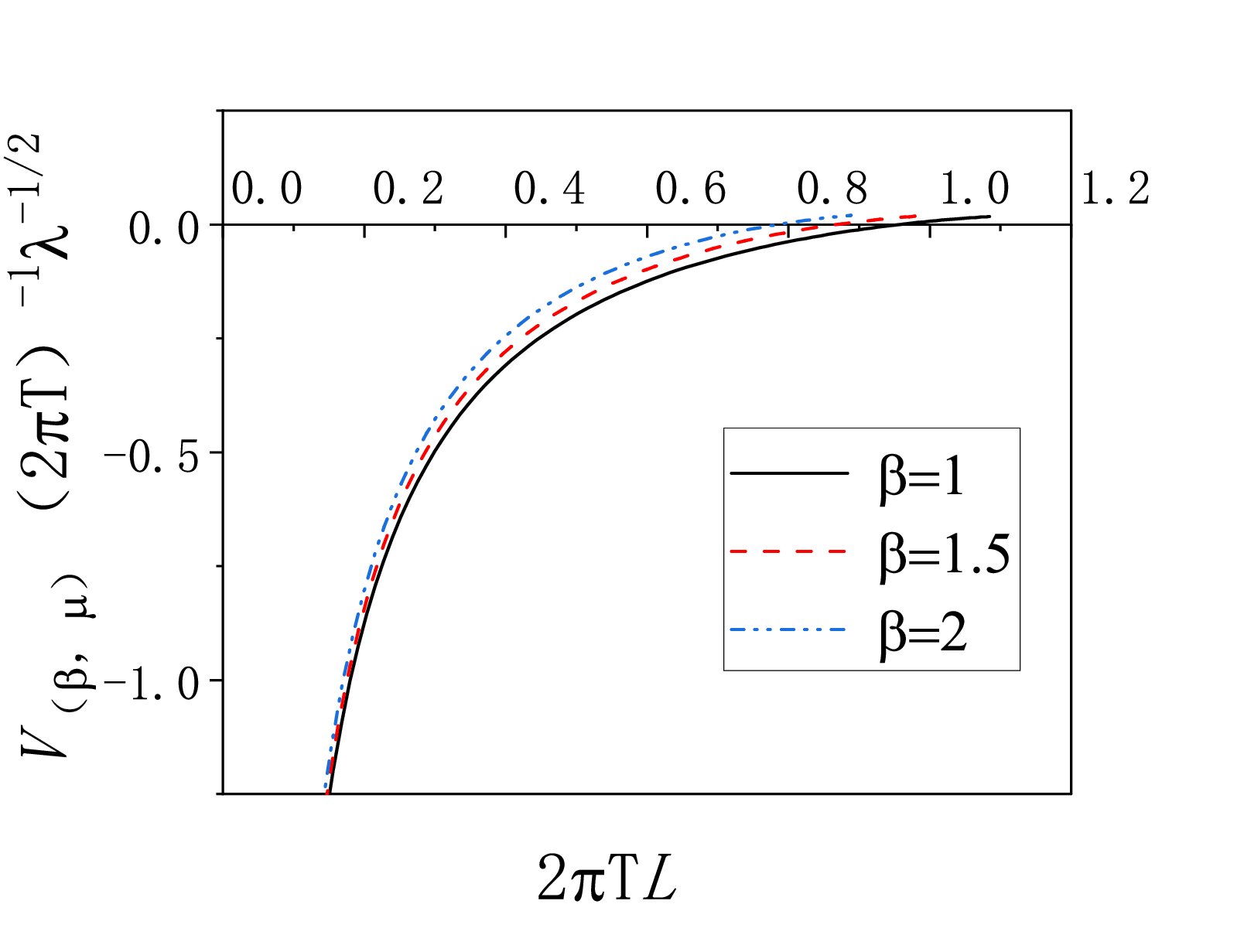}
\caption{Heavy quark potential at finite temperature. Left:
$\frac{V_{(\beta,\mu)}}{2\pi T\sqrt{\lambda}}$ versus $2\pi TL$
for $\mu=1,1.5,2$ with fixed $\beta=1$. Right: $\beta=1,1.5,2$
with fixed $\mu=1$. All quantities are in the dimensionless.}
\label{fig:hq_potential}
\end{figure}

Fig.~\ref{fig:hq_potential} shows $\frac{V_{(\beta,\mu)}}{2\pi
T\sqrt{\lambda}}$ versus $2\pi TL$. Increasing either $\mu$ or
$\beta$ shifts the potential upward and weakens the attractive
interaction, but the $\mu$ dependence is much stronger, consistent
with $|\chi_\mu^{V}|>|\chi_\beta^{V}|$ in the scanned region. This
agrees with earlier charged background results: \cite{Ewerz2015}
computed the screening distance, free energy, and running coupling
in nonconformal models with finite chemical potential and found a
reduced screening distance, while \cite{CWX} showed that
increasing the black hole charge raises the potential and weakens
the heavy quark interaction. The feature relevant here is that
$\mu$ remains dominant when $\beta$ is simultaneously nonzero. The
screening length $L_s$ decreases monotonically with both
parameters, and the reduction from $\mu$ is considerably larger,
qualitatively consistent with lattice studies of color screening
at finite temperature \cite{Q1,Q2}. Many holographic studies have
established that background parameters can significantly alter
color screening and quarkonium stability
\cite{HP1,HP2,HP3,HP4,HP5,HP6,HP8,HP9,HP10,HP11,HP12,HP13,HP14};
our results are consistent with these trends while isolating the
two scales within the EMa model.

\subsection{Probe pair free energy: $\beta$ dependence}
\begin{figure}
\centering
\includegraphics[width=0.48\textwidth]{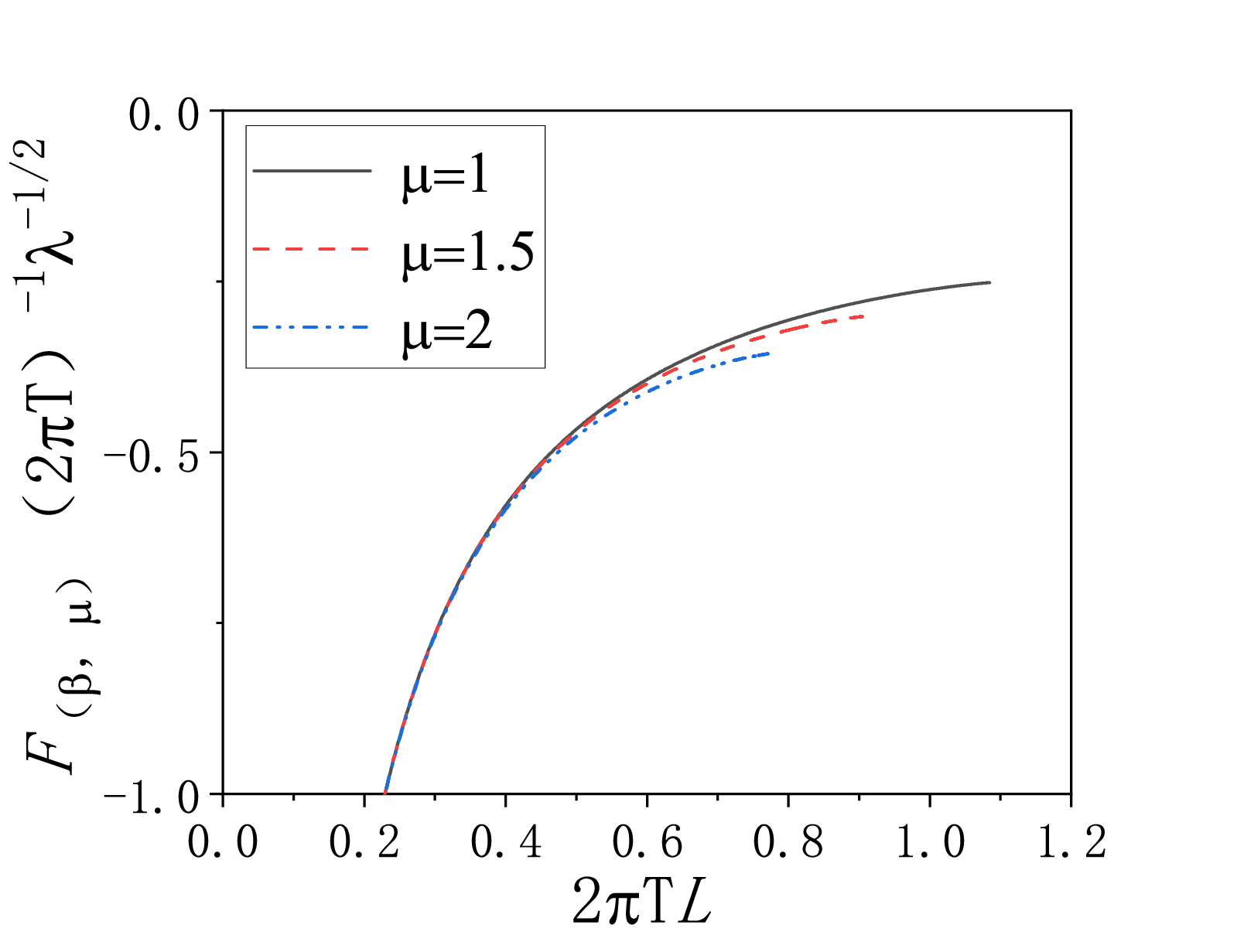}\hfill
\includegraphics[width=0.48\textwidth]{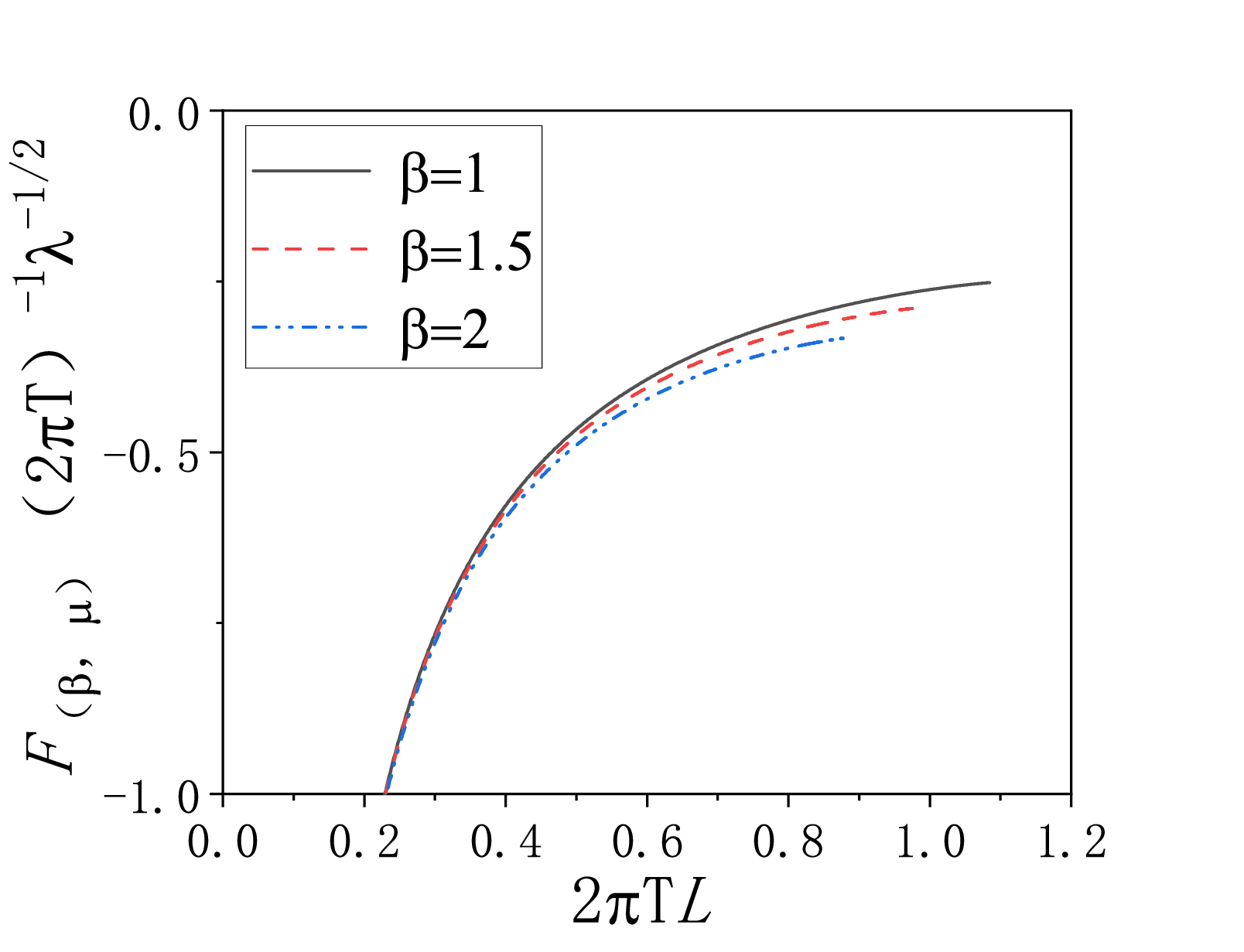}
\caption{Probe pair free energy at finite temperature. Left:
$\frac{F_{(\beta,\mu)}}{2\pi T\sqrt{\lambda}}$ versus $2\pi TL$
for $\mu=1,1.5,2$ with fixed $\beta=1$. Right: $\beta=1,1.5,2$
with fixed $\mu=1$. All quantities are in the dimensionless.}
\label{fig:free_energy}
\end{figure}

Fig.~\ref{fig:free_energy} shows $\frac{F_{(\beta,\mu)}}{2\pi
T\sqrt{\lambda}}$ versus $2\pi TL$. Unlike the potential, the
probe pair free energy is more sensitive to $\beta$ than to $\mu$
in the scanned region, i.e., $|\chi_\beta^{F}|>|\chi_\mu^{F}|$.
The free energy of the probe pair $F_{Q\bar{Q}}$ is distinct from
the background thermodynamic free energy. Attributing the observed
$\beta$ dependence directly to the horizon entropy $S$ of the
black brane through $F=U-TS$ would require equating the probe pair
free energy with the background free energy, which is not
justified. The probe pair entropy is $S_{Q\bar{Q}}=-\partial
F_{Q\bar{Q}}/\partial T$, which is not the same as $S$. A direct
thermodynamic decomposition of the observed $\beta$ dependence
would require computing $S_{Q\bar{Q}}$ and separating the
temperature derivatives. The present calculation establishes the
numerical hierarchy in the probe observable and its geometric
correlation with the constant $\beta^2$ term in $f(z)$, rather
than a thermodynamic causal attribution. Near dissociation, curves
at different $\mu$ diverge, consistent with higher density
shortening $L_s$, but the overall $\beta$ dependence of
$F_{Q\bar{Q}}$ is stronger in the scanned region. This is
consistent with the thermodynamic analysis of TSB backgrounds in
Refs.~\cite{TSBmajor,TSBmajor1} and with studies of dense
holographic matter in \cite{AH1,AH2,FS,DW1,SM}, though those
background results are not used to equate the two free energies.
We note that Rougemont et al.~\cite{Rougemont2016} constructed a
baryon-rich Einstein-Maxwell-Dilaton model calibrated to lattice
data and studied its thermodynamics and energy loss; their
framework does not isolate a separate momentum relaxation scale,
which is the additional control provided by the EMa geometry here.

\subsection{Running coupling}
\begin{figure}
\centering
\includegraphics[width=0.48\textwidth]{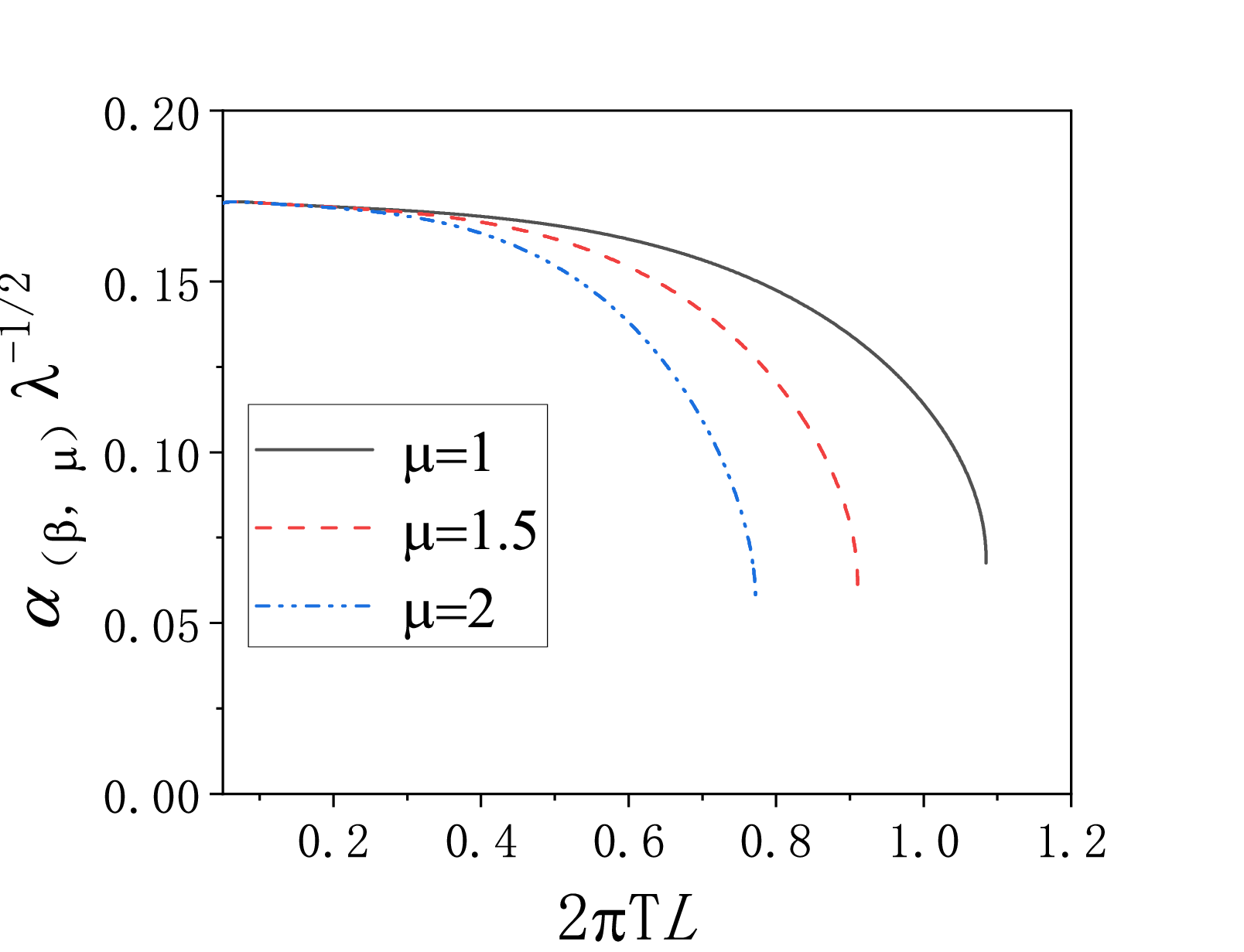}\hfill
\includegraphics[width=0.48\textwidth]{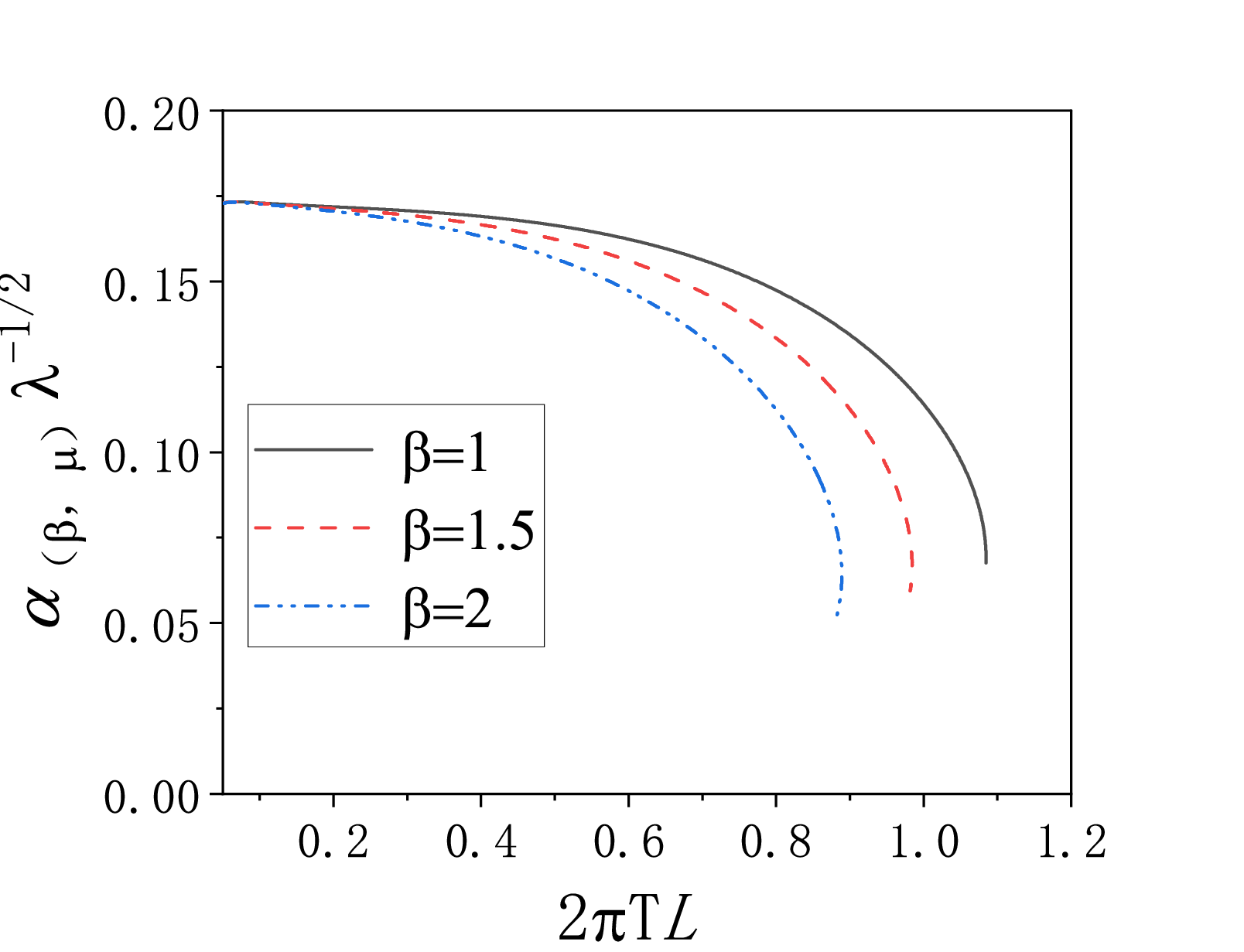}
\caption{Running coupling at finite temperature. Left:
$\frac{\alpha_{(\beta,\mu)}}{\sqrt{\lambda}}$ versus $2\pi TL$ for
$\mu=1,1.5,2$ with fixed $\beta=1$. Right: $\beta=1,1.5,2$ with
fixed $\mu=1$. All quantities are in the dimensionless.}
\label{fig:running_coupling}
\end{figure}

Fig.~\ref{fig:running_coupling} shows
$\frac{\alpha_{(\beta,\mu)}}{\sqrt{\lambda}}$ versus $2\pi TL$. As
with the potential, $\mu$ is the dominant suppressor. This is
qualitatively consistent with the charged background result of
\cite{hc10}, who found that increasing black hole charge reduces
the running coupling of the $Q\bar{Q}$ pair, indicating weakened
short range strong interactions. The magnetic field study of
Aref'eva et al.~\cite{Arefeva2024RC} concerns a different control
parameter, a strong magnetic field acting on light quarks, and is
not directly comparable to the $\mu$ and $\beta$ competition
considered here; we cite it only as an example of running coupling
suppression by an external scale. Related studies of the running
coupling in medium can be found in
\cite{DC,CX2,hc9,hc11,hc12,hc13,hc14,hc15}. In the present
background, $\frac{\alpha}{\sqrt{\lambda}}$ approaches a finite
plateau as $2\pi TL\to0$ and drops rapidly beyond $L_s$,
independent of whether $\mu$ or $\beta$ is varied.

\subsection{Geometric origin of the hierarchy}
The two parameters enter the metric function $f(z)$ differently.
The chemical potential contributes a term $\mu^2 z^4/(3z_h^4)$,
which grows toward the horizon and produces a finite Maxwell flux
in the near horizon region; this directly deforms the geometry
that controls the string turning point $z_c$ and therefore the
screening of static color interactions. In contrast, the TSB
parameter contributes a constant term $-\beta^2/4$, which is
independent of $z$ and shifts the horizon condition and the
overall geometry, but does not produce a localized near horizon
deformation of comparable strength. As a consequence, the
potential and running coupling track $\mu$ more strongly, while
the probe pair free energy tracks $\beta$ more strongly in the
scanned region. This separation persists across the scanned
parameter region and is not an artifact of the particular
parameter values shown in the figures, making the hierarchy a
useful model level diagnostic. It is not a proof of universality,
as a dilaton or non-Abelian flavor sector could modify the
relation between the geometry and the observables.

\section{Zero temperature limit}\label{sec:zeroT}

In this section, all quantities at zero temperature are expressed
in units set by the appropriate powers of $z_h$. At $T=0$, the
horizon condition becomes

\begin{equation}
\frac{1}{z_h^2}=\frac{\beta^2}{8}+\frac{\mu^2}{6}. \label{Th}
\end{equation}

Thus $z_h$ remains finite whenever $\mu$ and/or $\beta$ are
nonzero. Consequently, the Bekenstein-Hawking entropy density,
which scales as $1/z_h^3$, does not vanish at $T=0$; this model
has a finite ground state entropy, a property known for such
backgrounds \cite{TSBmajor1}. The temperature dependent part of
the free energy vanishes, but the finite horizon leaves a nonzero
difference between the potential and the probe pair free energy.
From Eqs.~\eqref{V0} and \eqref{Fc}, or directly from
Eq.~\eqref{VFrelation}, one has

\begin{equation}
\frac{F_{(\beta,\mu)}^{T=0}}{\sqrt{\lambda}}
=\frac{V_{(\beta,\mu)}^{T=0}}{\sqrt{\lambda}} -\frac{1}{\pi z_h},
\label{VT0}
\end{equation}
with $z_h$ given by Eq.~\eqref{Th}. Therefore $V^{T=0}\neq
F^{T=0}$ for all cases analyzed here. The running coupling is
defined from the free energy as
\begin{equation}
\frac{\alpha^{T=0}_{(\beta,\mu)}}{\sqrt{\lambda}} =\frac{3}{4}L^2
\frac{d\left(F^{T=0}_{(\beta,\mu)}/\sqrt{\lambda}\right)}{dL}.
\label{alphaT0}
\end{equation}

We separate the two pure cases and use the corrected relation
\eqref{VT0} whenever the free energy is discussed.

\subsection{$\beta=0$, finite $\mu$}

\begin{figure}
\centering
\includegraphics[width=0.48\textwidth]{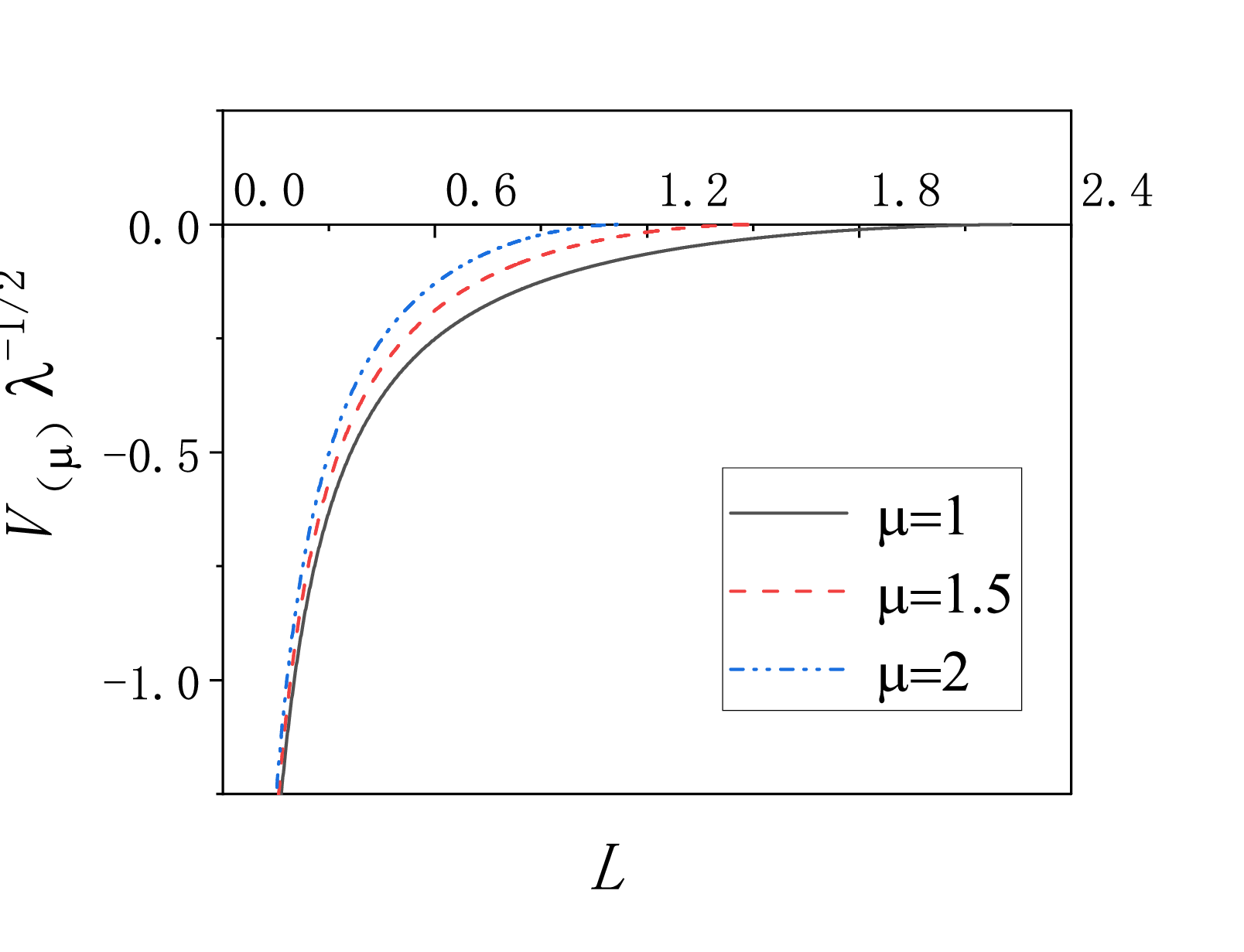}\hfill
\includegraphics[width=0.48\textwidth]{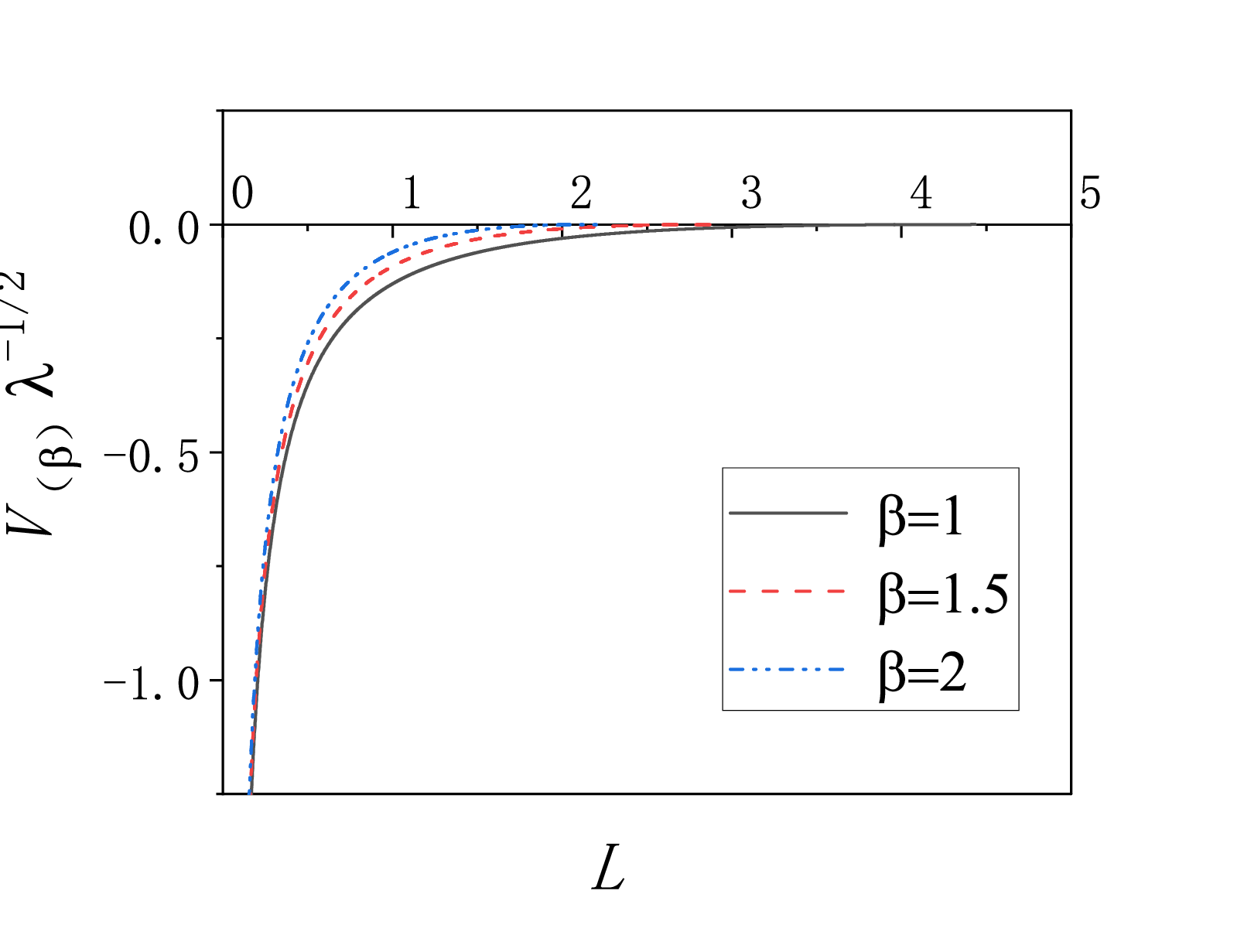}
\caption{Heavy quark potential at zero temperature. Left:
$V_{(\mu)}/\sqrt{\lambda}$ versus $L$ for $\mu=1,1.5,2$ with fixed
$\beta=0$. Right: $V_{(\beta)}/\sqrt{\lambda}$ versus $L$ for
$\beta=1,1.5,2$ with fixed $\mu=0$. All quantities are in units
set by the appropriate powers of $z_h$.}
\label{fig:zeroT_potential}
\end{figure}

\begin{figure}
\centering
\includegraphics[width=0.48\textwidth]{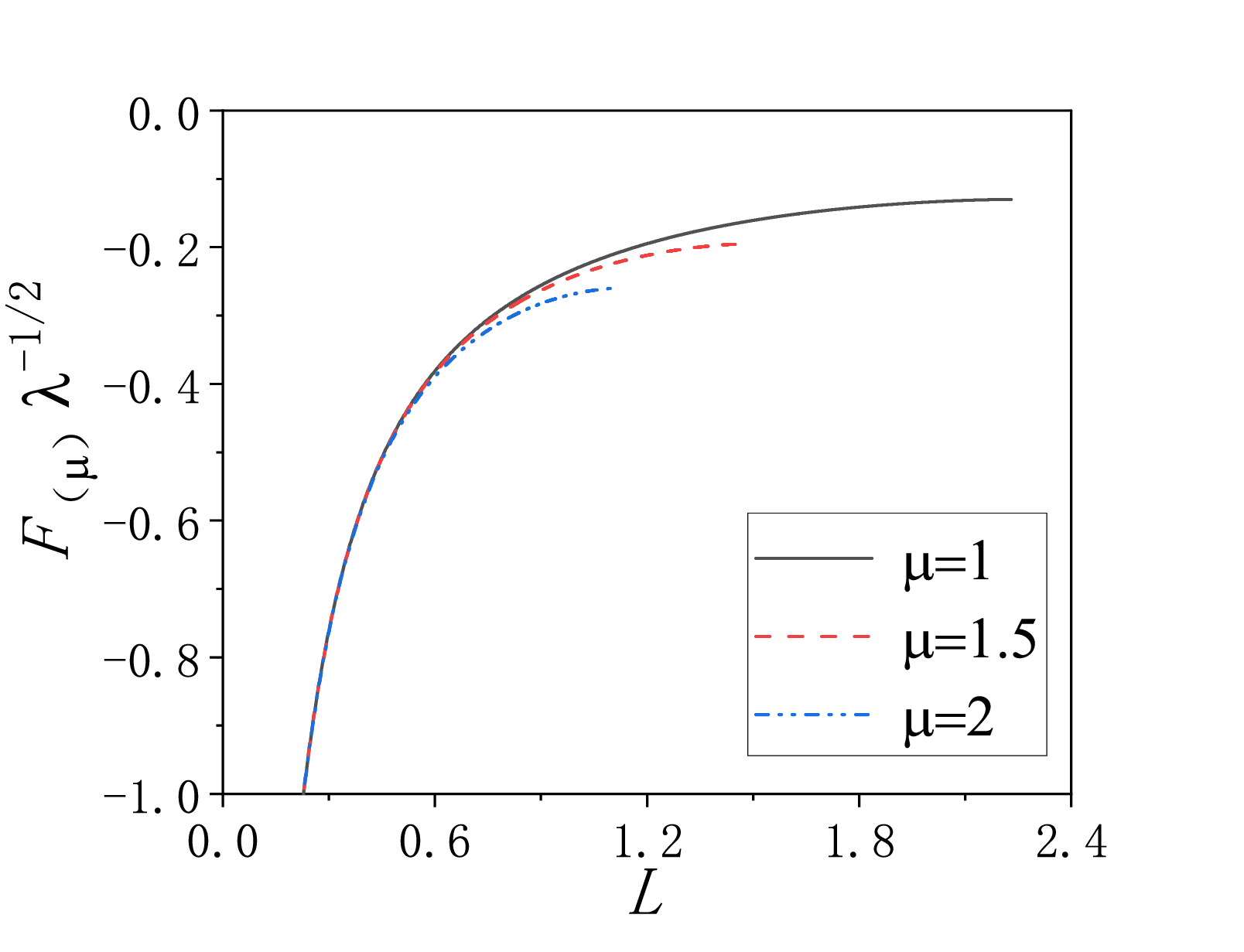}\hfill
\includegraphics[width=0.48\textwidth]{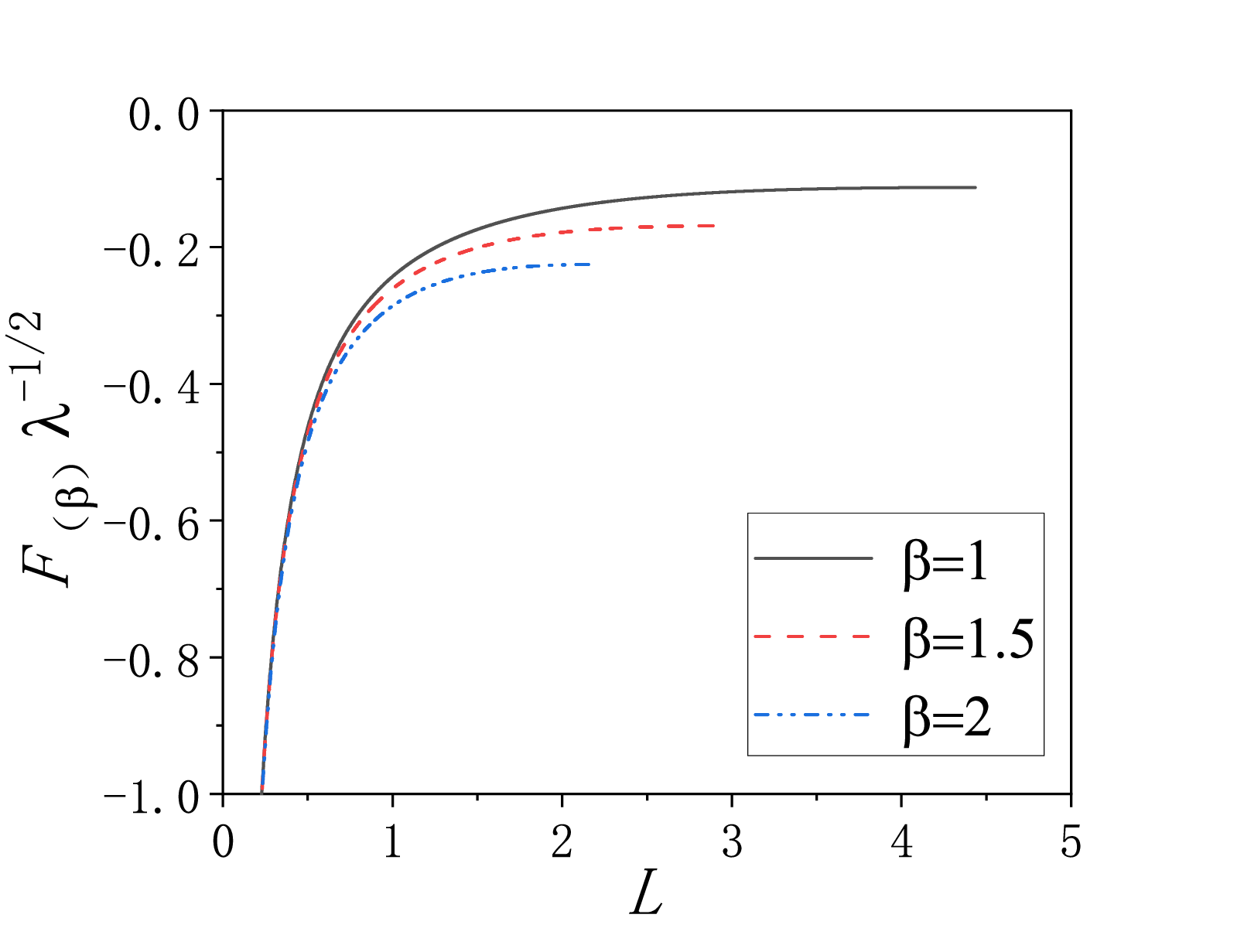}
\caption{Heavy quark free energy at zero temperature. Left:
$F_{(\mu)}/\sqrt{\lambda}$ versus $L$ for $\mu=1,1.5,2$ with fixed
$\beta=0$. Right: $F_{(\beta)}/\sqrt{\lambda}$ versus $L$ for
$\beta=1,1.5,2$ with fixed $\mu=0$. All quantities are in units
set by the appropriate powers of $z_h$.}
\label{fig:zeroT_free_energy}
\end{figure}

\begin{figure}
\centering
\includegraphics[width=0.48\textwidth]{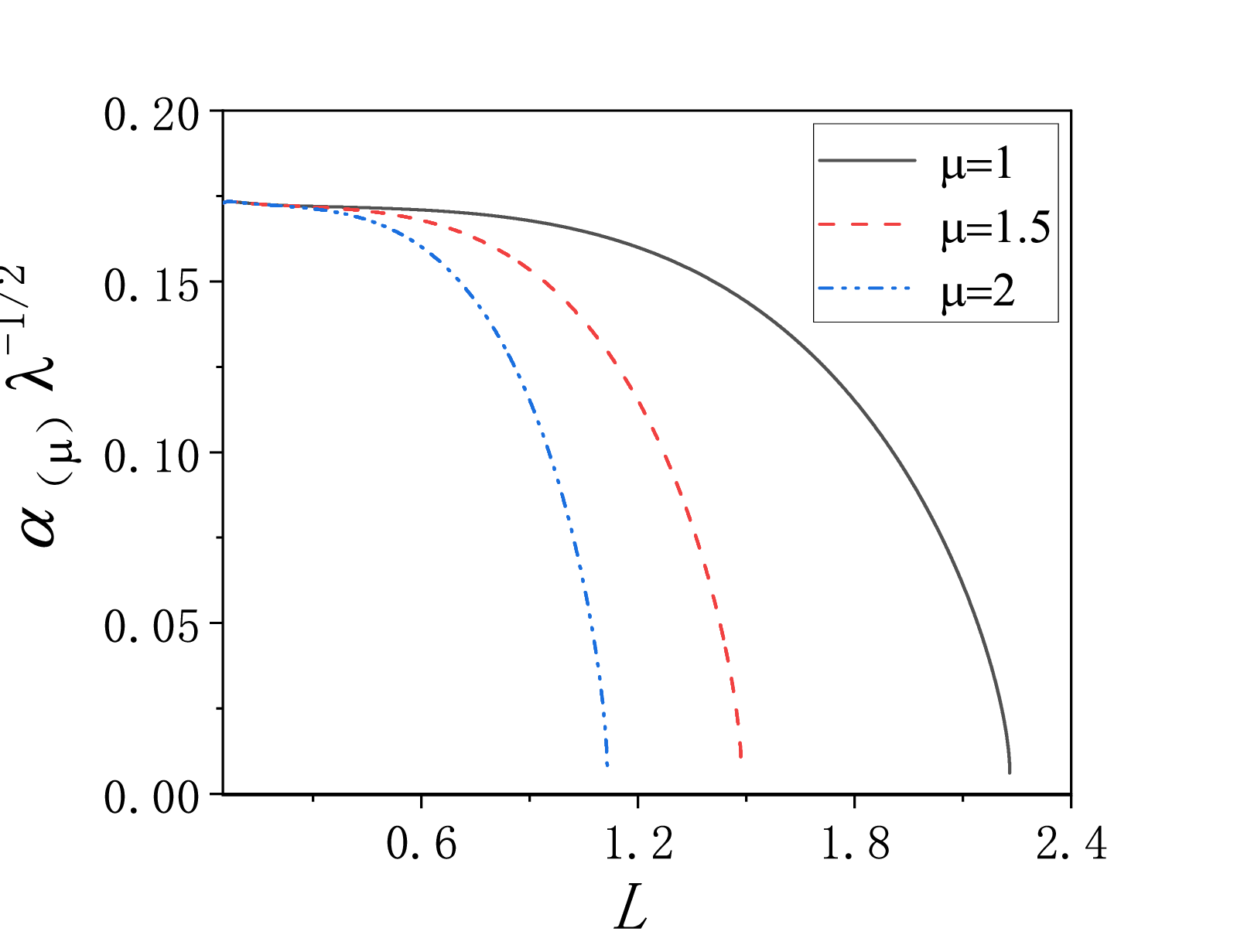}\hfill
\includegraphics[width=0.48\textwidth]{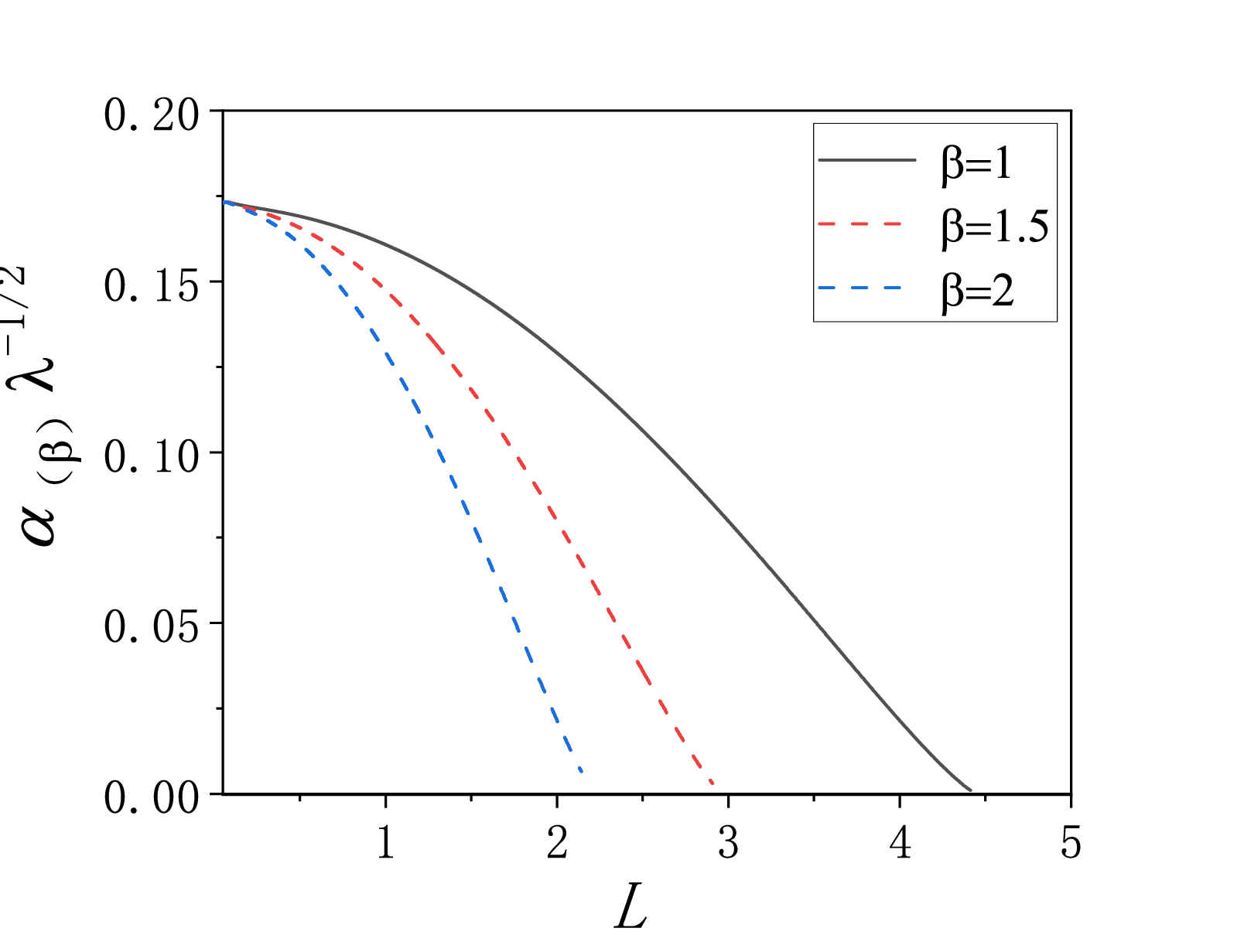}
\caption{Running coupling at zero temperature. Left:
$\alpha_{(\mu)}/\sqrt{\lambda}$ versus $L$ for $\mu=1,1.5,2$ with
fixed $\beta=0$. Right: $\alpha_{(\beta)}/\sqrt{\lambda}$ versus
$L$ for $\beta=1,1.5,2$ with fixed $\mu=0$. All quantities are in
units set by the appropriate powers of $z_h$.}
\label{fig:zeroT_running_coupling}
\end{figure}

As shown in the left panels of Figs.~\ref{fig:zeroT_potential},
\ref{fig:zeroT_free_energy}, and \ref{fig:zeroT_running_coupling},
the potential remains approximately Coulombic for all $\mu$, but
increasing $\mu$ raises it. Screening is weaker than at finite
$T$, and the dissociation distance moves to larger $L$. The zero
temperature density effect alone still shortens the dissociation
scale relative to the pure TSB case below. Meanwhile, increasing
$\mu$ decreases the free energy, but the effect is much weaker
than that on the potential. Increasing $\mu$ also suppresses the
running coupling. The horizon entropy density remains finite
because $z_h$ is finite from Eq.~\eqref{Th}; the relation in
Eq.~\eqref{VT0} shows that the probe pair free energy differs from
the potential by a term that depends on $\mu$ through $z_h$. The
variation of this term partially compensates for the increase in
the potential, resulting in a weaker $\mu$ dependence of the free
energy. These results are consistent with the zero temperature
screening analysis of Ref.~\cite{Ewerz2015} and with lattice
studies of the screening radius \cite{R1,R6}, within the model
level scope of this work.

\subsection{$\mu=0$, finite $\beta$}

As shown in the right panels of Figs.~\ref{fig:zeroT_potential},
\ref{fig:zeroT_free_energy}, and \ref{fig:zeroT_running_coupling},
with only momentum relaxation, larger $\beta$ also raises the
potential, but the effect is much milder. The dissociation
distance is nearly a factor of two larger than in the pure density
case, so pure TSB is inefficient at dissolving quarkonium in this
model. Meanwhile, increasing $\beta$ decreases the free energy
more significantly than increasing $\mu$. Compared with the finite
temperature results, the dominance of $\mu$ in the potential and
that of $\beta$ in the free energy appear more pronounced at zero
temperature. Increasing $\beta$ also suppresses the running
coupling, but the effect is much milder. A distinct feature is the
large $L$ behavior of $\alpha$: instead of a sharp drop, it
decreases slowly and approximately linearly, reflecting the smooth
radial modification of the metric by the axions. This qualitative
difference in the running coupling is a clear zero temperature
signature of the mechanism separation. The horizon remains finite
from Eq.~\eqref{Th}, so the entropy density does not vanish and
$V^{T=0}\neq F^{T=0}$; the extra term in Eq.~\eqref{VT0} depends
on $\beta$ through the horizon position $z_h$. Heavy quark
observables in purely momentum relaxing backgrounds have been
discussed in
\cite{TSB2,TSB3,TSB4,qx1,qx2,qx3,qx4,qx5,qx6,qx7,qx8,qx9,qx10,qx11};
the present comparison shows that these effects remain subleading
to density effects in the scanned region when both scales are
present. The asymmetry between the two cases at $T=0$ has the same
geometric origin as at finite temperature: the $\mu^2 z^4$ term is
concentrated near the horizon and deforms the region that controls
string screening, whereas the constant $\beta^2$ term shifts the
horizon globally without a comparable localized deformation.

\section{Conclusion and discussion}\label{sec:summary}

The main finding of this work is a scale-dependent hierarchy in
the EMa background. At fixed temperature, the density parameter
$\mu$ controls the static potential and the running coupling more
strongly than the momentum relaxation parameter $\beta$, whereas
$\beta$ controls the probe pair free energy more strongly. At
$T=0$ this hierarchy survives in the dissociation scale, and the
contrast between the two parameters becomes more pronounced. The
origin is geometric: the $\mu^2 z^4$ term grows toward the horizon
and deforms the region that sets the string turning point, while
the constant $\beta^2$ term shifts the entire geometry more
uniformly. The finite difference logarithmic sensitivities defined
in Eq.~\eqref{chi} make this statement quantitative; the
inequalities \eqref{ineq} hold throughout the scanned admissible
region, and the mixed logarithmic finite differences remain
subleading there.

This separation has a direct physical reading. Static color
screening is governed by the near horizon region where the string
turning point sits, and that region is most sensitive to the
density deformation. The probe pair free energy involves the full
string profile from the boundary to the turning point and is
therefore more sensitive to global changes of the geometry, which
the $\beta^2$ term provides. In a medium that is both dense and
dissipative, different quarkonium observables may thus respond to
different effective scales. Once both scales are present, no
single medium temperature or density can simultaneously capture
the potential, the free energy, and the running coupling.

The comparison also clarifies the relation to earlier studies.
Refs.~\cite{Ewerz2015,hc10} isolated density effects with
$\beta=0$, while
\cite{TSB2,TSB3,TSB4,qx1,qx2,qx3,qx4,qx5,qx6,qx7,qx8,qx9,qx10,qx11}
isolated momentum relaxation with $\mu=0$. Each of these studies
captures one limit correctly, but neither can determine which
scale dominates a given observable when both are present. The EMa
model provides a setting where this question has a definite answer
in the scanned region, and the answer is not uniform across
observables. Rougemont et al.~\cite{Rougemont2016} calibrated a
baryon-rich Einstein-Maxwell-Dilaton model to lattice data and
studied its thermodynamics and energy loss, but their framework
does not include an independent momentum relaxation scale. The
present comparison is complementary: it is not quantitatively
calibrated, but it isolates a mechanism that those studies cannot
address.

The limitations of the EMa model are important to state precisely.
It has no dilaton and no non-Abelian flavor sector, so its
equation of state does not match lattice QCD and its Abelian $\mu$
is not a baryon, charge, or strangeness chemical potential. At
$T=0$ the horizon remains finite, the entropy density does not
vanish, and the potential and probe pair free energy differ by
$1/(\pi\alpha' z_h)$. These features are specific to this
bottom-up geometry. The numerical values of $L_s$ and the
dissociation scale should not be compared directly with QCD. What
is more robust is the geometric separation: the $\mu$ term is
localized near the horizon, the $\beta$ term is not, and this
difference produces the observed hierarchy.

A natural next step is to repeat the comparison in a background
with a dilaton potential that reproduces the lattice equation of
state at $\mu=0$, and with a non-Abelian sector that gives the
$\mu$ term a clearer flavor interpretation. If the hierarchy
persists, it would suggest that the geometric localization of the
density deformation, rather than the specific axion realization,
is responsible for the pattern. If it does not, the ordering of
the sensitivities would indicate that the matter sector or the
symmetry breaking mechanism changes the relation between geometry
and observables in a qualitative way. Anisotropic TSB backgrounds
and moving quark observables would provide additional tests. The
finite difference logarithmic sensitivities introduced here are
convenient for such comparisons because they are dimensionless and
can be applied without recalibrating the model.

In summary, the EMa comparison shows that density and momentum
relaxation need not act on quarkonium observables through a single
common scale. The potential and running coupling track the near
horizon density deformation, while the probe pair free energy
tracks the global effect of momentum relaxation. This pattern is a
model level diagnostic: it is not a QCD prediction, but it
identifies a mechanism that can be tested, confirmed, or ruled out
in more realistic holographic constructions.

\section{Acknowledgments}\label{sec:ack}
This work is supported by the National Natural Science Foundation
of China (NSFC) under Grant Number~12375140 and the Fundamental
Research Funds for National Universities, China University of
Geosciences under Grant Number~2025XLB102.

\end{document}